\documentclass[conference]{IEEEtran}
\IEEEoverridecommandlockouts
\usepackage{cite}
\usepackage{amsmath,amssymb,amsfonts}
\usepackage{algorithmic}
\usepackage{graphicx}
\usepackage{textcomp}
\usepackage{caption}

\usepackage{soul,xcolor}
\usepackage{filecontents}
\usepackage[normalem]{ulem}

\graphicspath{{figures/}} 

\begin{document}
\title{Energy-Efficient and Delay-Guaranteed Joint Resource Allocation and DU Selection in O-RAN {
\thanks{This work is supported by Ontario Centers of Excellence
(OCE) 5G ENCQOR program and Ciena.}
}}
\author{\IEEEauthorblockN{Turgay Pamuklu, \IEEEmembership{Member, IEEE}, Shahram Mollahasani, \IEEEmembership{Member, IEEE}, Melike Erol-Kantarci,\\ \IEEEmembership{Senior Member, IEEE}
\{turgay.pamuklu, smollah2, melike.erolkantarci\}@uottawa.ca}
}
\maketitle
\makeatletter
\def\ps@IEEEtitlepagestyle{%
  \def\@oddfoot{\mycopyrightnotice}%
  \def\@oddhead{\hbox{}\@IEEEheaderstyle\leftmark\hfil\thepage}\relax
  \def\@evenhead{\@IEEEheaderstyle\thepage\hfil\leftmark\hbox{}}\relax
  \def\@evenfoot{}%
}
\def\mycopyrightnotice{%
  \begin{minipage}{\textwidth}
  \centering \scriptsize
Accepted Paper (DOI: 10.1109/5GWF52925.2021.00025). IEEE policy provides that authors are free to follow funder public access mandates to post accepted articles in repositories. When posting in a repository, the IEEE embargo period is 24 months. However, IEEE recognizes that posting requirements and embargo periods vary by funder. IEEE authors may comply with requirements to deposit their accepted manuscripts in a repository per funder requirements where the embargo is less than 24 months.
  \end{minipage}
}
\makeatother

\begin{abstract}
The radio access network (RAN) part of the next-generation wireless networks will require efficient solutions for satisfying low latency and high-throughput services. The open RAN (O-RAN) is one of the candidates to achieve this goal, in addition to increasing vendor diversity and promoting openness. In the O-RAN architecture, network functions are executed in central units (CU), distributed units (DU), and radio units (RU). These entities are virtualized on general-purpose CPUs and form a processing pool. These processing pools can be located in different geographical places and have limited capacity, affecting the energy consumption and the performance of networks. Additionally, since user demand is not deterministic, special attention should be paid to allocating resource blocks to users by ensuring their expected quality of service for latency-sensitive traffic flows. In this paper, we propose a joint optimization solution to enhance energy efficiency and provide delay guarantees to the users in the O-RAN architecture. We formulate this novel problem and linearize it to provide a solution with a mixed-integer linear problem (MILP) solver. We compare this with a baseline that addresses this optimization problem using a disjoint approach. The results show that our approach outperforms the baseline method in terms of energy efficiency.

\end{abstract}


\section{Introduction}
\begin{figure*}[t]
\centering
\includegraphics[width=0.7\textwidth]{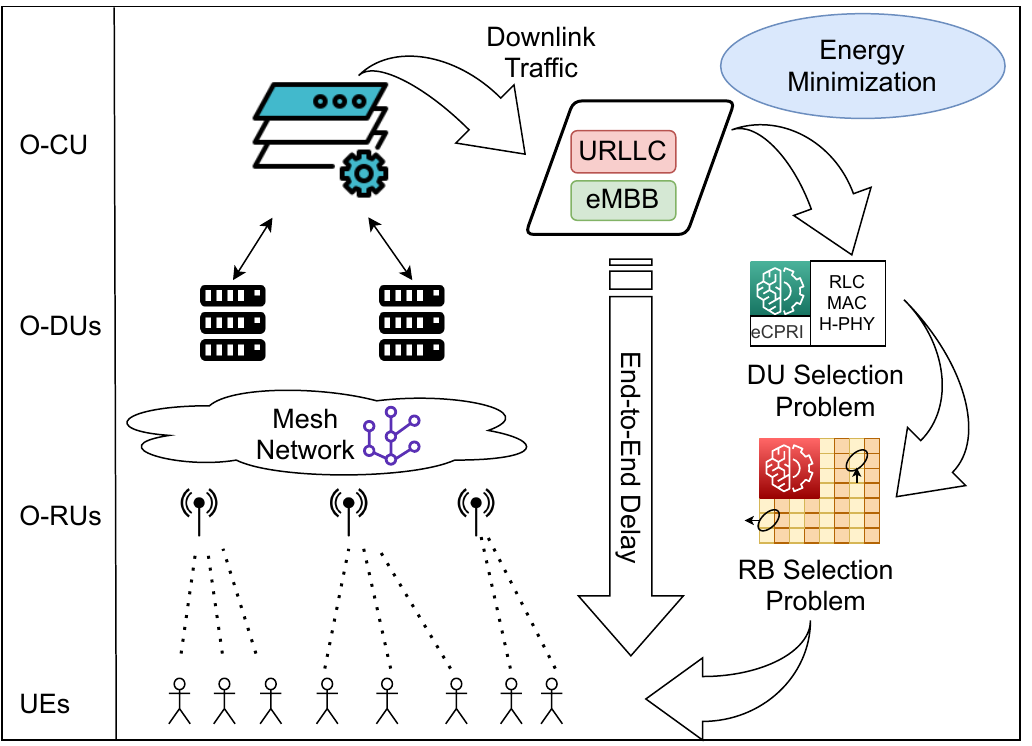}
\caption{\label{fig:arch} Overall architecture of the joint DU and RB selection problem in a disaggregated O-RAN. All RUs have physical connections to all DUs forming a mesh connected architecture. (O-CU: open center unit, O-DU: open distributed unit, O-RU: open radio unit, RLC: radio link control, MAC: medium access control, PHY: high physical layer, eCPRI: enhanced common public radio interface, RB: resource block, URLLC: Ultra-reliable low-latency communication, eMBB: enhanced mobile broadband).}
\end{figure*}

\par Disaggregation of user and control plane functions, along with the openness of interfaces, are the two main properties of Open RAN (O-RAN) that grant vast flexibility for vendor selection, and consequently, lead to improved key performance indicators (KPIs) \cite{O-RANAlliance2020}. Energy efficiency is one of the fundamental KPIs whose importance is more and more emphasized as the operational expenditure costs of Telcos are boosting, and the concerns around global warming are increasing \cite{Pamuklu2020}. Furthermore, the growth in mobile traffic elevates these costs more than ever \cite{Pawar2020}. Therefore, one of the main objectives of the proposed solutions for new RAN architectures should be reducing the overall energy consumption.
\par End-to-end delay is another essential KPI which especially becomes important for ultra-reliable low-latency communication (URLLC) traffic \cite{Elsayed2019}.  Meanwhile, satisfying the throughput requirement of enhanced mobile broadband (eMBB) is also equally important. The overall delay that each packet experiences is affected by the scheduling delay, and it should be kept below the predefined delay budget for the traffic type of the user \cite{Mollahasani2021}. Additionally, due to the disaggregation of network entities in O-RAN, user traffic experiences fronthaul propagation delay during data transfer between the RU and the DU. Therefore, optimizing the resource block (RB) allocation and choosing proper DUs with respect to their locations is critical to balancing the energy consumption and satisfying a delay guarantee in the network.

\par RB selection is a well-studied problem in 5G. Elsayed et al. focus on machine learning (ML) approaches for RB selection in a network with different traffic types \cite{Elsayed2019}. In another study, they improve their ML algorithms by including transfer reinforcement learning in their model. As a result, their distributed approach enhances the performance and scalability of their solutions \cite{Elsayed2020}. Dominic et al. propose a distributed solution for the RB selection problem in which the interest is in device-to-device (D2D) communications \cite{Dominic2020}. 
\par Additionally, in recent works, RB allocation models are updated based on the New Radio (NR) features of 5G. For example, Korrai et al., by considering mixed-numerologies in 5G networks, tackle the RB allocation problem \cite{Korrai2020}. Their technique reduces energy consumption in case of imperfect channel state information. Abiko et al. investigate the RB allocation problem for a network slicing scenario, and they propose a scalable distributed reinforcement learning approach \cite{Abiko2020}. Their solution dynamically changes the resource allocations to the slices according to the total number of slices in the network.
\par Disaggregation approaches yield a new location and mapping selection problem between different tiers in the next-generation wireless network architectures. For example, Mouwad et al. deal with baseband unit and remote radio head mapping problems in the Cloud RAN architecture \cite{Mouawad2019}. Furthermore, virtual network function (VNF) and network slicing technologies boost the need for novel location selection algorithms. Attaoui et al. propose an end-to-end delay minimization model for the VNF placement problem in the radio access network \cite{Attaoui2020}. In our prior work, we have proposed a cost-efficient green Radio model for O-RAN by considering a mesh network where all tiers are connected, and the operational expenditure (OPEX) is minimized by making optimum use of renewable energy  \cite{Pamuklu2020}. Nevertheless, this study does not address the RB allocation problem. Lastly, in another prior work, we combine the DU selection problem with the RB allocation problem, in which two nested reinforcement learning algorithms are used \cite{Mollahasani2021a}. However, this study does not focus on delay constraints.

\par In this study, we combine RB allocation and DU selection problems to enhance energy efficiency and guarantee delay for low-latency traffic. Specifically, we propose an energy-aware optimization model and jointly perform RB allocation and DU selection in the O-RAN architecture. We then linearize this problem to solve it with a mixed-integer linear problem (MILP) solver. We compare this scheme to a baseline which introduces a solution to the problems of RB allocation and DU selection, disjointly. Our results show that a joint method can significantly reduce the energy consumption in the O-RAN architecture.

The rest of the article is organized as follows: Section II explains the system model and describes the energy consumption minimization problem. Section III provides the computational experiments, and in Section IV, we conclude the paper.

\section{System Model}
\par Figure~\ref{fig:arch} shows the overall model of the joint RB and DU selection problem in the downlink for the O-RAN architecture \cite{O-RANAlliance2020}.\footnote{While we demonstrate the downlink traffic, our method can be easily generalized for the uplink.} Let $R$ be the number of RUs in the network. Each of these RUs ($r\in\mathcal{R}$) decides how to allocate physical RBs (PRBs) to $I$ user equipments (UEs) ($i\in\mathcal{I}$). The RUs are connected to $L$ DUs ($l\in\mathcal{L}$) with a mesh topology. Thus, RUs have the flexibility to choose a DU to operate their higher level network functions. Meanwhile, UEs demand $K$ number of traffic types ($k\in\mathcal{K}$) which have different data sizes ($U_{itk}$) and delay budgets ($\delta_{k}$). Thus, RUs should make traffic-aware allocation decisions to improve the performance of the network.

\par In this work, we focus on two critical KPIs. The first one is the overall energy consumption in the network, which grows into critical levels by increasing data traffic demands and the number of UEs in O-RAN \cite{Pamuklu2021}. The second one is the end-to-end delay which is significant for ultra-reliable low-latency communication (URLLC) traffic. DU and RB selection are two critical problems that have remarkable impacts on energy consumption and end-to-end delay, respectively. Our proposed solution aims to tackle these problems jointly. 

\begin{table}
\centering
\caption{\label{tab:Notations} Summary of the notations}
\begin{tabular}{c|p{6cm}}
\textbf{Sets} & \textbf{Explanation} \\ \hline
$r\in\mathcal{R}$ & set of RBs in one TTI\\
$t\in\mathcal{T}$ & set of TTIs \\
$i\in\mathcal{I}$ & set of UEs \\
$j\in\mathcal{J}$ & set of RUs \\
$l\in\mathcal{L}$ & set of DUs \\
$k\in\mathcal{K}$ & type of traffic \\ \hline
\textbf{Variables}& \textbf{Explanation} \\ \hline
$a_{itjrt'}$ & RB $rt'$ in RU $j$ is selected for user demand $it$ \\
$b_{jlt}$ & DU $l$ is selected for RU $j$ in time interval $t$  \\
$c_{lt}$ & DU $l$ is active in time interval $t$ \\
$u_{itk}$ & $k$ type traffic demanded by user $i$ in $t$ \\
$y_{itjt'}$ & at least one RB in RU $j$ is selected in time $t'$ for the demand $it$ \\ \hline
\textbf{Given Data}& \textbf{Explanation} \\ \hline
$U_{itk}$ & data size of $k$ type traffic demanded by user $i$ in $t$ \\
$S_{jirt}$ & max. data rate between RU $j$ and user $i$ for RB $rt$ \\
$D_{jl}^P$ & propagation delay between RU $j$ and DU $l$ \\
$\delta_{k}$ & delay budget of traffic type $k$ \\
$E_{l}^S$ & fixed energy consumption in DU $l$ \\
$E_{l}^D$ & dynamic energy consumption in DU $l$ \\
$TTI$ & transmission time interval  \\
$\mathcal{M}$ & Big M, a very large number  \\
\hline
\end{tabular}
\end{table}

\subsection{Proposed Energy Minimization Problem with a Guaranteed Maximum Delay Scheme}
To address the joint RB \& DU selection problem and reduce energy consumption with a guaranteed delay, we propose a MILP based problem formulation. Table~\ref{tab:Notations} summarizes the notations which are used in the problem formulation. Our objective function (Eq.~\ref{eq:obj}) is a summation of two main energy consumption factors in the DUs. The first one is fixed energy consumption, $E^{S}_{l}$, which can be reduced by switching off the DUs ($c_{lt'}$).  Note that a DU can be switched off only in a case that if that DU does not serve any RUs in the network. The second one is dynamic energy consumption $E^{D}_{l}$ which depends on the DU selection ($b_{jlt'}$) and RB selection ($a_{itjrt'}$) decisions.  
While the percentage of dynamic energy consumption is remarkably low in the overall energy consumption, Telcos may consider to take this into account for larger networks.

\par Our first constraint (Eq.~\ref{eq:delay}) handles our latency KPI by providing a guaranteed end-to-end delay for different types of traffic. As seen from the equation, there are two primary sources of latency. The first one is the scheduling delay because of the RB selection problem, which equals to the time difference between when a request arrives from a user traffic ($t$) until the assignment of the last RB to this request ($\max(y_{itjt'}*t')$). The second term comes from the propagation delay ($D_{jl}^{P}$) between RUs and DUs. Here, $\delta_{k}$ is the delay budget which depends on the type of traffic demand coming from the user. The second constraint (Eq.~\ref{eq:QoS}) satisfies the user demands in RUs. The data rate of an RU $S_{ijrt'}$ should be sufficient to serve their associated UEs ($a_{itjrt'}$). The $u_{itk}$ binary decision value is set up from the user demand ($U_{itk}$) to linearize this equation for different traffic types as formulated in eq.~\ref{eq:Uandu}. If the user demand is higher than zero, $u_{itk}$ is forced to be equal to one.
\par Eq.~\ref{eq:RBOneandOnly} ensures each RB is assigned to only one user.  Eq.~\ref{eq:RUOneandOnly} maintains the UEs-RUs mappings. In our network model, a UE demand may be satisfied by only one RU in a time interval. Eq.~\ref{eq:DUOneandOnly} provides that one RU can be assigned to only one DU. If an RU serves to any user in a time interval ($t'$), we have to bind that RU to one of the DU in this time interval. Eq~\ref{eq:DUOneandOnly2} guarantees this condition. Therefore users can get their expected network function services from the assigned DU. Eq.~\ref{eq:YandA} generates a correlation between two binary decisions related to the RB selection. Thus, we can reduce the complexity of our delay calculation and improve the performance of our solution by using $y_{itjt'}$ instead of $a_{itjrt'}$. Eq.~\ref{eq:DUswitch} imposes to activate a DU if there is an RU in the network which is delegated to that DU.  Finally, Eq.~\ref{eq:trafficOneandOnly} restricts user demand with one traffic type in a specific time interval.
\par This joint selection problem can be solved with a MILP solver in addition to heuristics \cite{Fendt2018}. In the next section, we analyze the solver results by comparing them with the results of a traditional disjoint approach.

\begin{align}
&\textbf{Minimize:}& \notag
\\
&\sum\limits_{\substack{t'\in\mathcal{T} \\ l\in\mathcal{L}}} \left[E^{S}_{l}* c_{lt'} + \sum\limits_{t=0}^{t'}\sum\limits_{\substack{{i\in\mathcal{I}} \\ {j\in\mathcal{J}} \\ r\in\mathcal{R}}}  E^{D}_{l} * b_{jlt'} * a_{itjrt'}  \right]
\label{eq:obj}
\\
&\textbf{Subject to:} \notag
\\
&\left(\max\limits_{\substack{j\in\mathcal{J} \\ {t'\in\mathcal{T}}}} (y_{itjt'}*t')-t\right) * TTI +\sum\limits_{\substack{{l\in\mathcal{L}} \\ {j\in\mathcal{J}} \\r\in\mathcal{R} \\ {t'\in\mathcal{T}}}} a_{itjrt'} * b_{jlt'} * D_{jl}^{P} \notag \\
&\leq \delta_{k} * u_{itk} , \;\quad\qquad\qquad\qquad \forall i\in\mathcal{I}, \forall t\in\mathcal{T},\forall k\in\mathcal{K}
\label{eq:delay} 
\\
&\sum\limits_{t'=t}^{\mathcal{T}}\sum\limits_{j\in\mathcal{J}} \sum\limits_{r\in\mathcal{R}}  S_{ijrt'} * a_{itjrt'} \geq \sum\limits_{k\in\mathcal{K}} (U_{itk} * u_{itk}), \notag \\
&\;\;\;\qquad\qquad\qquad\qquad\qquad\qquad\qquad\qquad \forall i\in\mathcal{I}, \forall t\in\mathcal{T}
\label{eq:QoS}
\\
&\mathcal{M} * u_{itk} - U_{itk} \geq 0, \;\;\qquad\quad\forall i\in\mathcal{I},\forall t\in\mathcal{T}, \forall k\in\mathcal{K}
\label{eq:Uandu} 
\\
&\sum\limits_{i\in\mathcal{I}}\sum\limits_{t\in\mathcal{T}} a_{itjrt'} \leq 1,\;\qquad\qquad\forall j\in\mathcal{J}, \forall r\in\mathcal{R},\forall t'\in\mathcal{T}
\label{eq:RBOneandOnly}
\\
&\sum\limits_{j\in\mathcal{J}} a_{itjrt'} \leq 1,\qquad\quad\forall i\in\mathcal{I},\forall t\in\mathcal{T}, \forall r\in\mathcal{R},\forall t'\in\mathcal{T}
\label{eq:RUOneandOnly}
\\
&\sum\limits_{l\in\mathcal{L}} b_{jlt'} \leq 1,\;\qquad\qquad\qquad\qquad\qquad\forall j\in\mathcal{J},\forall t'\in\mathcal{T}
\label{eq:DUOneandOnly}
\\
&\mathcal{M} * \sum\limits_{l\in\mathcal{L}} b_{jlt'} - \sum\limits_{\substack{{i\in\mathcal{I}} \\ {t\in\mathcal{T}}}} y_{itjt'} \geq 0,\;\qquad\quad\forall j\in\mathcal{J}, \forall t'\in\mathcal{T}
\label{eq:DUOneandOnly2} 
\\
&\mathcal{M} * y_{itjt'} -\sum\limits_{r\in\mathcal{R}} a_{itjrt'} \geq 0, \notag \\
&\;\;\qquad\qquad\qquad\qquad\quad\forall i\in\mathcal{I},\forall t\in\mathcal{T}, \forall j\in\mathcal{J}, \forall t'\in\mathcal{T}
\label{eq:YandA} 
\\
&\mathcal{M} * c_{lt} - \sum\limits_{j\in\mathcal{J}} b_{jlt} \geq 0,\;\qquad\qquad\qquad\forall l\in\mathcal{L},\forall t\in\mathcal{T}
\label{eq:DUswitch} 
\\
&\sum\limits_{k\in\mathcal{K}}  u_{itk} \leq 1, \;\;\qquad\qquad\qquad\qquad\qquad\forall i\in\mathcal{I},\forall t\in\mathcal{T}
\label{eq:trafficOneandOnly} 
\end{align}

\section{Computational Experiments}
\par
\begin{figure}
\centering
\includegraphics[width=0.45\textwidth]{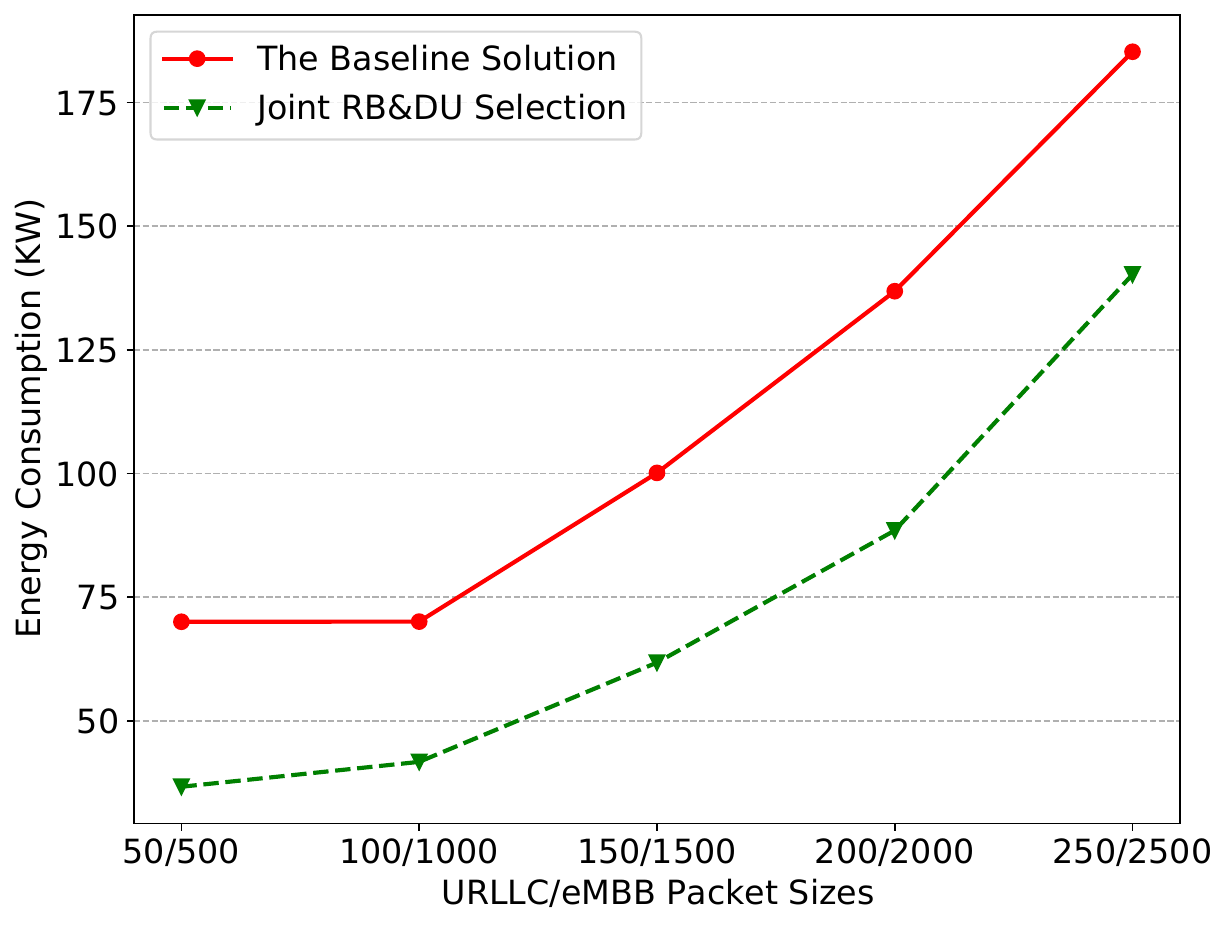}
\caption{\label{fig:obj} Total energy consumption (objective function) for varying packet sizes. Baseline is the disjoint solution.}
\end{figure}

 We implement simulations with three DUs ($L=3$) that serve six RUs ($R=6$) in an O-RAN network. The total number of users in the network is twelve ($I=12$), and each user demands two types of traffic ($K=2$). The first traffic type is URLLC, in which packet size is $U_{it0} = 50$ bits,\footnote{Note that this packet size is used in the lowest traffic rate scenario. We multiply each packet size from one to five to see the impact of the traffic rates in Figure~\ref{fig:obj}.} and the delay budget is $\delta_{0}$ = 2 ms. The other traffic type is eMBB with $U_{it1} = 500$ bits packet size and a $\delta_{1}=10$ ms delay budget. The arrival rate of eMBB traffic is three times higher than the URLLC traffic. We simulate the system for ten time intervals ($T=10$). In the first six time intervals, the arriving of a packet is independent for each user and are equally likely (uniformly distributed) with a 66\% average arriving rate.
\par The other parameters used in the simulations are shown in Table~\ref{tab:Comp}. The propagation delay $D_{jl}^P$ equals zero if an RU selects the nearest DU; otherwise, for all other DUs $D_{jl}^P$ is set to 2 ms. As a MILP solver, we use GUROBI to solve this energy minimization problem \cite{GurobiOptimization2021}. In this solver, we limit the simulations to twelve hours and a $10\%$ lower bound gap to compare with the baseline more accurately. The simulation results are based on the average of 6 runs with different seed values.
\begin{table}
\centering
\caption{\label{tab:Comp} Computational Parameters.}
\begin{tabular}{c|c|c}
Instance & Unit & Value \\ \hline
$S_{jirt}$ & bits & 350 \\
$D_{jl}^P$ & ms & [0,2] \\
$E_{l}^D$ & Wh & [1,1,1] \\
$TTI$ & ms & 1 \\
$\mathcal{M}$ & - & $10^{4}$ \\
\end{tabular}
\end{table}
\par The baseline is chosen as a method in which RB and DU selections are performed disjointly and concurrently in a MILP Solver. First, we use the same fixed energy consumption model for each DU ($E_{l}^S = 10kWh$). Figure~\ref{fig:obj} shows the energy consumption comparison of our proposed joint solution and the baseline for varying packet sizes. Although the baseline method provides a feasible solution without violating the end-to-end delay constraint, the energy consumption minimization performance reduces with large packet sizes. This outcome originates from RB selection decisions. If the network is underutilized, we can assign resource blocks immediately to the user demands in the same time interval. In case the packet size becomes more prominent than the serving data rates, RUs may not have enough RBs to allocate to UEs during that time interval. Therefore, it becomes critical to determine which user's demand needs to be addressed in the present time interval. Meanwhile, we can easily make this decision in the proposed joint solution by considering the propagation delay between RUs and DUs. Therefore, in the cases of the packet sizes larger than 150/1500 bits for the URLLC and eMBB traffic, the energy consumption of the proposed solution is at least 40\% lower than the baseline solution.
\begin{figure*}
\centering
\includegraphics[width=0.55\textwidth]{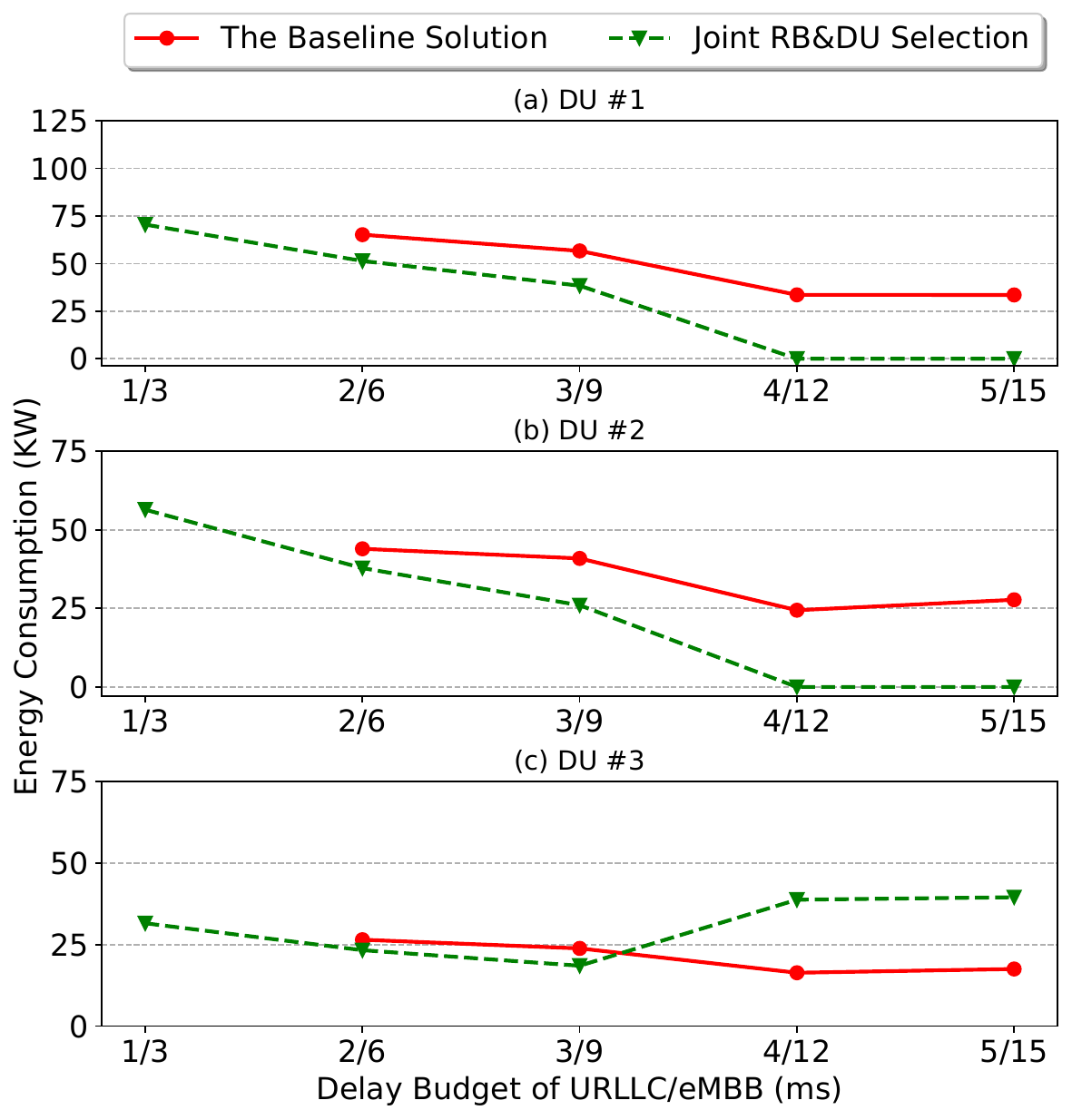}
\caption{\label{fig:delay} Impact of delay budget ($\delta_{k}$) variation on energy consumption for each DU. (a): First DU, (b): Second DU, (c): Third DU. }
\end{figure*}

\par Figure~\ref{fig:delay} details the performance improvement in scheduling and propagation delays when our method is used. In this figure, x-axis shows the delay budget instances for each traffic type. In this figure, we show the energy consumption of each DU separately in the Y-axis while the delay budget for URLLC  ($\delta_{0}$) changes from 1 ms to 5 ms in X-axis, and the delay budget for eMBB ($\delta_{1}$) is three times higher than the budget for URLLC (e.g., 1/3 means $\delta_{0}$=1 ms and $\delta_{1}$=3 ms delay budgets for URLLC and eMBB traffic, respectively.). Lastly, we use a heterogeneous fixed energy consumption model for DUs ($E_{l}^S = [15,10,5]$) to show the adaptation of our joint solution to reduce the overall energy consumption. As seen from the figure, the baseline solution could not find a feasible solution in the case that the delay budgets are equal to $\delta_{0}$=1 ms and $\delta_{1}$=3 ms for the URLLC and eMBB traffic types, respectively. Moreover, our joint model performs better with the higher delay budgets by migrating the functions to the third DU (Figure~\ref{fig:delay}c). The third DU has lower fixed energy consumption than the other DUs. For the sake of this migration, for instance, one can switch off the other higher energy consuming DUs, and thus, reduce the total energy consumption in the network. As seen from Figure~\ref{fig:delay}, the energy consumption of the third DU in our proposed approach is higher than the baseline because of load migration from other DUs.  Note that fairness between DUs is not aimed in this study, we are rather interested in reducing the network's overall energy consumption. On the other hand, the baseline model could not provide this migration efficiently due to delay constraint violation. For this reason, higher energy consuming DU1 and DU2 stay active and consume at least 25 kW energy in any case, whilst the proposed solution consumes zero energy in the 4/12 and 5/15 delay budget cases.  

\par
\begin{figure}
\centering
\includegraphics[width=0.45\textwidth]{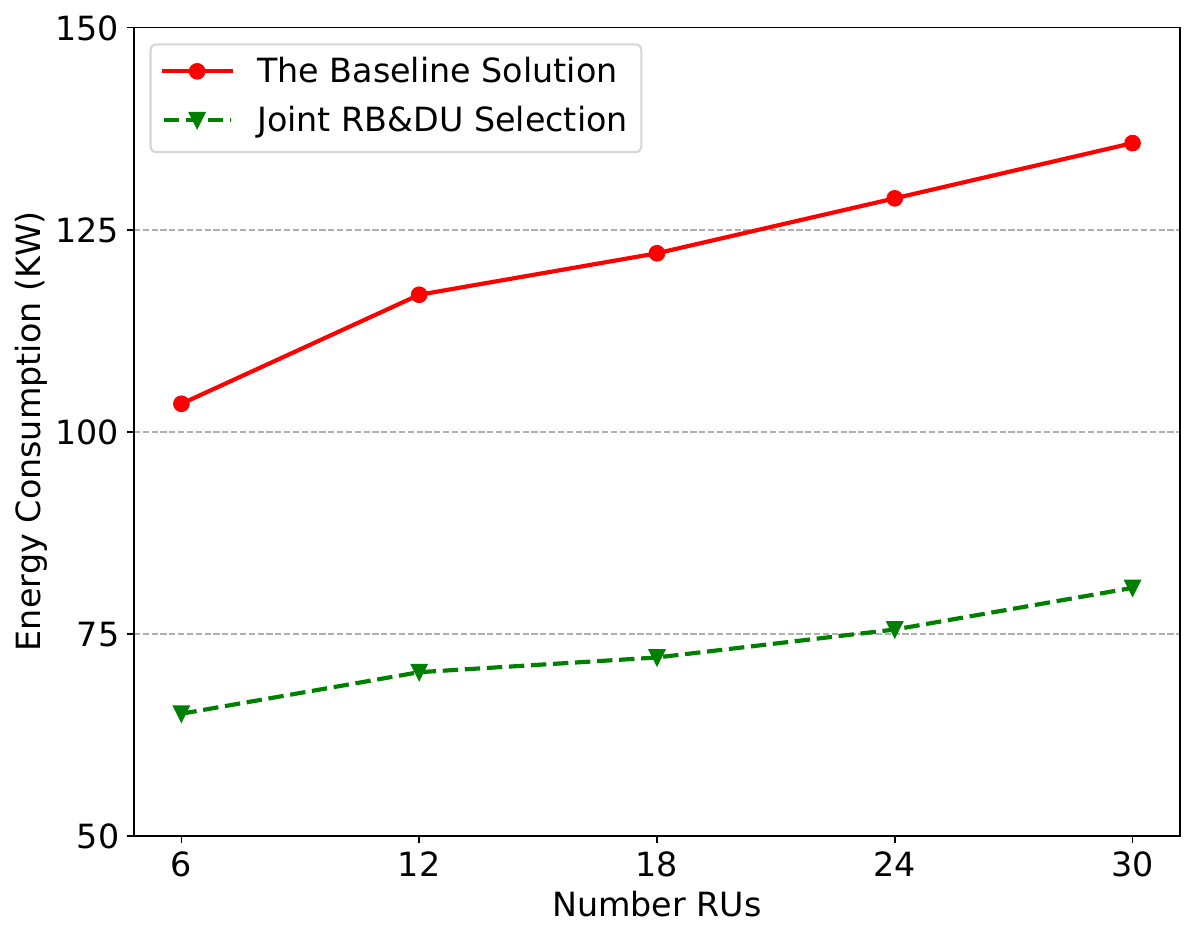}
\caption{\label{fig:support} Energy consumption for varying number of RUs.}
\end{figure}

 Finally, Figure~\ref{fig:support} shows how the joint solution performs with the change of network size. In this figure, we choose the packet size 150 and 1500 bits for URLLC and eMBB packets, respectively. Then, we increase the number of RUs in the network from 6 to 30 to see the performance of the proposed solution for larger networks. This figure demonstrates that the energy consumption is lower in our proposed joined solution even if a large number of RUs are present. For instance, in the case that the network has 30 RUs, our proposed solution incurs 40\% less energy consumption than the baseline solution. 

\section{Conclusion}
\par In this paper, we propose a solution that aims at two critical KPIs in O-RAN: energy minimization and limiting the latency for delay-sensitive user traffic. We focus on DU selection and RB allocation problems. We formulate the joint DU selection and RB allocation as a MILP problem. We compare our proposed solution to a disjoint DU selection and RB allocation scheme which is used as a baseline. Our results show that a joint selection approach is critical to minimize the energy consumption in the O-RAN architecture. The disjoint baseline approach is inefficient in migrating the DUs functions due to the strict budgets for each selection problem. Meanwhile, the balancing capability of our joint selection approach between the scheduling and propagation delay boosts the amount of migrations between DUs, and allows energy minimization. We show that our proposed approach also has lower energy consumption under a larger network size.

\bibliography{wf21}

\end{document}